\documentclass[prd,amsmath,amssymb]{revtex4}

\usepackage{graphicx}
\usepackage{dcolumn}
\usepackage{bm}
\usepackage{cases}
\usepackage{mathtools}
\DeclarePairedDelimiter\bra{\langle}{\rvert}
\DeclarePairedDelimiter\ket{\lvert}{\rangle}
\DeclarePairedDelimiterX\braket[2]{\langle}{\rangle}{#1\,\delimsize\vert\,\mathopen{}#2}
\usepackage{pgfplots}
\pgfplotsset{compat=1.8}
\usetikzlibrary{arrows.meta}
\usepackage{tikz}
\usetikzlibrary{quantikz}
\usepackage{listings} 

\begin{document}

\title{Quantum LLMs Using Quantum Computing to Analyze and Process Semantic Information}

\author{Timo\hspace*{1mm}Aukusti\hspace*{1mm}Laine \vspace*{0.5cm}}
\email{timo@financialphysicslab.com}

\begin{abstract}
We present a quantum computing approach to analyzing Large Language Model (LLM) embeddings, leveraging complex-valued representations and modeling semantic relationships using quantum mechanical principles. By establishing a direct mapping between LLM semantic spaces and quantum circuits, we demonstrate the feasibility of estimating semantic similarity using quantum hardware. One of the key results is the experimental calculation of cosine similarity between Google Sentence Transformer embeddings using a real quantum computer, providing a tangible demonstration of a quantum approach to semantic analysis. This work reveals a connection between LLMs and quantum mechanics, suggesting that these principles can offer new perspectives on semantic representation and processing, and paving the way for future development of quantum algorithms for natural language processing.
\end{abstract}

\maketitle

\section{Introduction}

Large Language Models (LLMs) have achieved remarkable success in natural language processing, yet their underlying mechanisms remain only partially understood. This paper explores a fundamental question: Can quantum mechanics, as a theoretical framework, provide valid and useful insights into the structure and behavior of LLMs? We hypothesize that quantum mechanics, with its inherent ability to represent complex relationships and probabilistic behavior, can offer a valuable lens for understanding and potentially manipulating LLM semantic spaces. This is particularly relevant given the increasing computational demands of LLMs, which motivates the exploration of alternative computational paradigms. Furthermore, we envision that quantum algorithms may offer significant advantages, such as the potential for exponential speedups in calculating semantic similarity within high-dimensional embedding spaces if more efficient quantum algorithms can be developed, or the ability to uncover subtle semantic relationships that are difficult to detect using classical methods. However, we acknowledge that current quantum hardware and algorithms face significant limitations, particularly in terms of qubit scaling and coherence.

Current LLMs primarily rely on real-valued embeddings, which may limit their capacity to capture subtle semantic nuances such as antonymy, sentiment, or relational information. To address this, we introduce an approach that incorporates complex-valued embeddings, leveraging principles from quantum mechanics to represent semantic information with both magnitude and phase. This framework establishes a direct mapping between LLM semantic spaces and quantum circuits, enabling the potential application of quantum algorithms to language-based tasks.
This work aims to address the following research questions:
1.  Can quantum mechanics, as a theoretical framework, provide valid and useful insights into the structure and behavior of LLMs, leading to a deeper understanding of their emergent properties?
2.  How accurately can quantum circuits compute semantic similarity between LLM embeddings compared to classical methods, providing concrete evidence for the applicability of quantum principles to semantic analysis?
3.  What are the key challenges and opportunities in utilizing current real quantum computers for natural language processing tasks involving LLM embeddings, considering the limitations of qubit availability and coherence?

One of the key contribution of this paper is the demonstration, using a real-world LLM (Google's Sentence Transformer), of a direct, experimentally verifiable connection between LLM embedding spaces and quantum circuits. We leverage this connection to explore real embedding spaces and calculate semantic similarity on a real quantum computer. We illustrate the principles of our approach by simulating the double-slit experiment within a quantum circuit, providing a tangible demonstration of quantum mechanics applied to semantic modeling. This provides evidence supporting the validity of a quantum-inspired approach to LLMs.

The paper is structured as follows: We begin by reviewing relevant background on LLMs, complex embeddings, and quantum mechanics. We then describe our approach for mapping LLM embeddings to quantum circuits and calculating semantic similarity. Next, we present the results of our experiments on a real quantum computer, comparing them with classical calculations. Finally, we discuss the implications of our findings, limitations, and future research directions, focusing on the broader potential of quantum mechanics as a framework for understanding and enhancing LLMs.

\section{Background and Related Work}

Large Language Models (LLMs) have revolutionized Natural Language Processing (NLP), achieving state-of-the-art results in various tasks such as text generation, translation, and question answering. At the heart of these models lies a sophisticated mechanism for representing text: high-dimensional vector embeddings. These embeddings map words, phrases, and even entire documents into a continuous semantic space, where geometric relationships reflect semantic similarities. The Transformer architecture \cite{ref_vaswani}, with its self-attention mechanisms, has been key to capturing long-range dependencies in text and enabling LLMs to scale to unprecedented sizes, as seen in models like BERT \cite{ref_devlin} and GPT-3 \cite{ref_brown}.

While these embedding spaces are often treated as continuous for practical purposes, a fundamental aspect of LLMs hints at an underlying discreteness: their reliance on a finite vocabulary of tokens. This discrete foundation suggests that the seemingly continuous semantic space might possess a quantized structure, analogous to the discrete energy levels observed in quantum systems. This inherent quantization prompts a compelling question: can we leverage the theoretical frameworks of mathematical physics and tools of quantum mechanics to gain a deeper understanding of the organization and dynamics of these semantic spaces, and potentially uncover previously hidden relationships?

Several approaches have been proposed for analyzing LLM embedding spaces. Dimensionality reduction techniques, such as Principal Component Analysis (PCA) \cite{ref_jolliffe} and t-distributed Stochastic Neighbor Embedding (t-SNE) \cite{ref_maaten}, are often used to visualize the embedding space in lower dimensions. Geometric analysis techniques, such as calculating cosine similarity between vectors \cite{ref_salton}, are used to quantify the semantic similarity between words and phrases. These approaches have provided valuable insights into the organization of semantic information in LLMs, but they often fail to capture more nuanced relationships such as antonymy, analogy, or hierarchical structures. The geometry of these embedding spaces has been explored by researchers like Mimno and Thompson \cite{ref_mimno}, while Tenney et al. \cite{ref_tenney} have investigated how context influences these embeddings. 

The field of quantum computing offers a different paradigm for computation, leveraging the principles of quantum mechanics to solve problems intractable for classical computers \cite{ref_nielsen}. Quantum circuits, composed of qubits and quantum gates, are used to implement quantum algorithms \cite{ref_divincenzo}. Key quantum algorithms, such as Shor's algorithm for factoring \cite{ref_shor} and Grover's algorithm for searching \cite{ref_grover}, demonstrate the potential for speedups in certain computational tasks. The development of quantum hardware is rapidly advancing, with increasing qubit counts and improved coherence times \cite{ref_arute}. However, the noisy intermediate-scale quantum (NISQ) era presents significant challenges for realizing the full potential of quantum algorithms \cite{ref_bharti}.

Quantum-inspired models have also found applications in cognitive science, offering a framework for representing concepts and relationships in a more flexible manner than traditional models. These models have been used to address phenomena such as contextuality and ambiguity in human cognition \citep{ref_bruza, ref_busemeyer,ref_atmanspacher,ref_khrennikov}. Approach utilizes quantum formalism to represent uncertainty and superposition in cognitive processes. Furthermore, the use of complex embeddings in machine learning, particularly in knowledge graph embedding models like ComplEx \citep{ref_trouillon}, RotatE \citep{ref_sun}, and QuatE \citep{ref_zhang}, provides a richer representation of data by encoding both magnitude and phase information.

More recently, the field of quantum machine learning (QML) has emerged, exploring the potential of quantum algorithms to enhance machine learning tasks. Variational Quantum Algorithms (VQAs) \cite{ref_bharti} are a prominent approach, using parameterized quantum circuits to approximate solutions to optimization problems. Quantum feature maps and kernels \cite{ref_havlicek} offer the potential to create feature spaces that are intractable for classical computers, potentially leading to improved machine learning performance. Quantum neural networks (QNNs) \cite{ref_beer} are also being investigated as a way to leverage quantum computation for neural network training and inference. However, the development of practical QML algorithms faces significant challenges, including the limitations of current quantum hardware and the difficulty of demonstrating a quantum advantage \cite{ref_arrasmith}.

In the realm of quantum natural language processing (QNLP), researchers are exploring ways to use quantum algorithms for NLP tasks. Quantum language models are being developed to represent and process language data in a quantum framework. Compositional QNLP \cite{ref_coecke} focuses on combining the meanings of words and phrases using quantum-inspired mathematical structures. While QNLP is still in its early stages, it holds the potential to unlock new capabilities in natural language understanding and processing.

Our work builds upon these existing approaches by exploring the analogy between LLM embedding spaces and quantum mechanics, with a focus on experimentally demonstrating the applicability of quantum principles. We extend the standard, real-valued embedding space to the complex domain to model semantic interference effects, drawing parallels to phenomena such as the double-slit experiment in quantum mechanics. We build upon the theoretical foundation laid by previous works by Laine \citep{ref_laine1, ref_laine2}, which introduced the concept of "semantic wave functions" to describe the representation of meaning in LLMs, drawing an analogy between words and sentences and quantum wave functions. This paper provides a concrete implementation of these ideas using quantum circuits, demonstrates the feasibility of calculating LLM embeddings on a real quantum computer, and explores qubit reduction using complex embeddings.

\section{LLM Embeddings}

This section provides a review of LLM embeddings and cosine similarity, concepts that are central to this paper.

LLM embeddings are vector representations of text designed to capture semantic meaning. They enable tasks such as semantic search and enhance downstream machine learning models by providing a richer and more compact representation of text compared to raw words. These embeddings are crucial for understanding and processing language beyond simple keyword matching.

In LLMs, embedding vectors are typically real-valued

\begin{align}
    {\bf E}_N(\text{text})&= [e_1, e_2, e_3, ..., e_{N}], \label{eq_real_embedding}
\end{align}

\noindent
where $e_i$ are real numbers and $N$ is the dimension of the embedding space. The relative similarity between two embedding vectors is often quantified using cosine similarity, a technique, for example, employed in sentence transformers and Retrieval-Augmented Generation (RAG) implementations for retrieving semantically similar context or the best-matching chunks from a vector database.

Given two real-valued vectors ${\bf E}_1$ and ${\bf E}_2$, their inner product is defined as

\begin{equation}
   {\bf E}_1 \cdot {\bf E}_2 =  ||{\bf E}_1 || \ ||{\bf E}_2|| \cos\theta, \label{eq_inner_product}
\end{equation}

\noindent
which leads to the definition of cosine similarity

\begin{equation}
    S_C({\bf E}_1,{\bf E}_2)= \cos\theta = \frac{{\bf E}_1\cdot {\bf E}_2}{||{\bf E}_1|| \ ||{\bf E}_2||}. \label{eq_cos_sim_real}
\end{equation}

\noindent
Cosine similarity is a normalized measure, with values ranging from -1 to 1. A value of 1 indicates perfectly aligned (highly similar) vectors, while -1 signifies the most dissimilar vectors. In real-valued vector embedding spaces, cosine similarity provides a valuable measure of the semantic similarity between words, sentences, or documents, assessing similarity based on subject matter, independent of the length of the documents.

\subsection{Complex Embeddings}

This paper explores complex representations, specifically complex number embeddings. Complex embeddings are vector representations that utilize complex numbers to encode both magnitude and phase, offering a richer data representation compared to real-valued embeddings. They are particularly effective in modeling relationships and capturing nuanced information, especially in knowledge graphs and relational reasoning tasks. While providing increased representational capacity, they also introduce computational complexity relative to real-valued embeddings. Complex embeddings are defined as

\begin{align}
    {\bf E}_{CN}&= [e_{c1}e^{i\varphi_1}, e_{c2}e^{i\varphi_2}, e_{c3}e^{i\varphi_3}, ..., e_{cN}e^{i\varphi_N}], \label{eq_complex_embedding}
\end{align}

\noindent
where $e_{ci}$ is a real number and $\varphi_i$ is an angle of rotation. Real-valued embedding vectors can be derived from complex vectors by setting $\varphi_i=0$

\begin{equation}
  e_i = e_{ci}e^{i\varphi_i=0} \label{eq_complex_transform}
\end{equation}

\noindent
For complex vectors, denoted as ${\bf a}$ and ${\bf b}$, the complex cosine similarity is defined as

\begin{equation}
    {\bf S}_{C}({\bf a},{\bf b})= \frac{{\bf a}^H {\bf b}}{||\mathbf{a}|| \ ||\mathbf{b}||}, \label{eq_cos_sim_complex}
\end{equation}

\noindent
where ${\bf a}^H$ denotes the conjugate transpose (Hermitian transpose) of the vector ${\bf a}$. The complex cosine similarity can be a real number, but it is generally a complex number.

The relative importance of the phase and magnitude components of the complex cosine similarity depends on the specific application and the definition of semantic similarity. In some cases, the magnitude is dominant, indicating that phase is less significant. This occurs when alignment strength is paramount, where the magnitude, $|{\bf S}_C(\mathbf{a}, \mathbf{b})|$, reflects the strength of alignment, irrespective of relative phase. Phase-invariant similarity also falls into this category, where vectors are considered similar despite phase shifts, emphasizing the magnitude. Similarly, in shape or pattern recognition, the overall shape or pattern is more important than absolute phase, making the magnitude more relevant.

Conversely, the phase component can be of primary importance. This is evident in relative timing or synchronization problems, such as with signals or oscillations, where relative timing is crucial, and the phase reflects the phase difference. Interference phenomena in physics also highlight the importance of phase, as phase differences determine interference patterns. Furthermore, when information is encoded in signal phase, phase comparison is essential, as is the case with phase-encoded information. If complex numbers represent directions, the phase encodes the angle, making it a crucial factor.

\section{Double-Slit Experiment}

\begin{figure}[h!]
    \centering
    \includegraphics[width=0.8\textwidth]{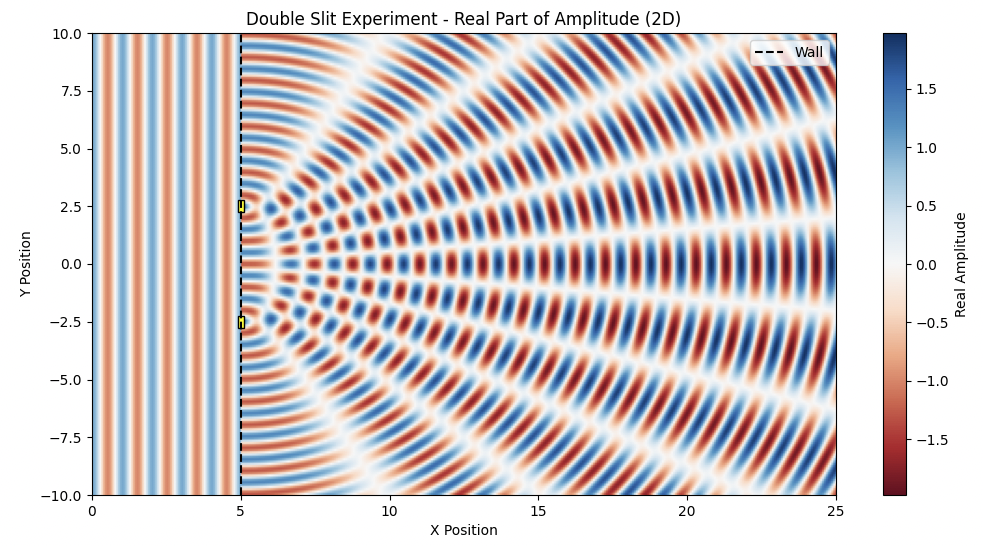}
\caption{Real-valued amplitude distribution in the double-slit experiment. This serves as an analogy for the representation of semantic information in traditional, real-valued LLM embeddings.}
    \label{fig:real-amplitude}
\end{figure}

\begin{figure}[h!]
    \centering
    \includegraphics[width=0.8\textwidth]{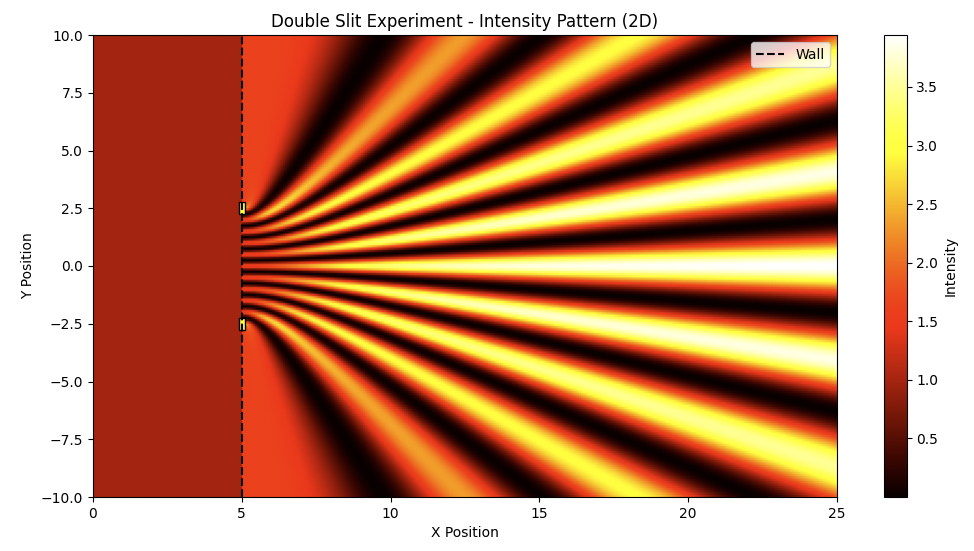}
\caption{Intensity distribution on the detection screen in the double-slit experiment, showing the alternating regions of high and low intensity resulting from wave interference.}
    \label{fig:two-slit}
\end{figure}

To illustrate the concepts of embedding spaces and complex phases, this paper utilizes the double-slit experiment as an analogy. As discussed in reference \cite{ref_laine1}, the double-slit experiment provides a framework for understanding potential semantic interference effects. The key aspects are illustrated in Figures \ref{fig:real-amplitude} and \ref{fig:two-slit}.

The classic double-slit experiment involves directing a stream of particles (e.g., photons or electrons) towards a barrier with two openings. The resulting pattern observed on a detection screen exhibits an interference pattern, a characteristic signature of wave behavior. Instead of a simple sum of probabilities from each slit acting independently, the probability distribution reveals alternating regions of high and low intensity, known as interference fringes. This phenomenon arises from the superposition, or combination, of the waves emanating from each slit.

Mathematically, the wave functions describing particles passing through each slit can be expressed as

\begin{align}
    \psi_1(x,y) &= A(x,y) e^{i\varphi}, \\
    \psi_2(x,y) &= B(x,y) e^{i\phi},
\end{align}

\noindent
where $A(x,y)$ and $B(x,y)$ represent the amplitudes of the waves at a specific location $(x, y)$ on the detection screen, and $\varphi$ and $\phi$ represent their respective phases. The probability of detecting a particle at that location is then given by

\begin{equation}
    P(x,y) =  |\psi_1(x,y) + \psi_2(x,y)|^2 = 
    |A(x,y)|^2 + |B(x,y)|^2 + 2 |A(x,y)| |B(x,y)| \cos(\varphi - \phi). \label{eq_interference}
\end{equation}

\noindent
The term $2 |A(x,y)| |B(x,y)| \cos(\varphi - \phi)$, known as the interference term, is crucial. It arises from the superposition of complex probability amplitudes. A description relying solely on real numbers would fail to capture the phase relationships essential for explaining the interference pattern. The double-slit experiment demonstrates the wave-particle duality inherent in light and matter. The phase difference $\varphi - \phi$ is key to the interference pattern. In the context of Large Language Models, each "slit" can be interpreted as representing a distinct semantic context, and the wave function represents the word, sentence, or paragraph being analyzed. The interference pattern then reflects the potential semantic relationships between these contexts.

\subsection{Double-Slit Experiment in a Quantum Circuit}

This section explores a quantum circuit model of the double-slit experiment. Quantum circuits are composed of qubits and quantum gates. A qubit, the fundamental unit of quantum information, exists in a superposition of two distinct states, denoted as $\ket{0}$ and $\ket{1}$. The Hadamard gate is a single-qubit quantum gate that transforms a definite state into a superposition of equally probable $\ket{0}$ and $\ket{1}$ states. The Hadamard gate is used in quantum algorithms and is important for quantum computational advantage. In the context of the double-slit experiment, qubits can represent the possible paths of a particle, and the Hadamard gate can be used to create a superposition of these paths, analogous to the particle existing in both slits simultaneously.

To illustrate how interference manifests within a quantum circuit, consider a qubit with basis states $\ket{0}$ and $\ket{1}$. We begin with an arbitrary initial complex state

\begin{align}
    \ket{\psi}_{initial} &= \psi_1\ket{0} + \psi_2\ket{1} = \alpha_ 1 e^{i\varphi_1} \ket{0} + \beta_1  e^{i\phi_1} \ket{1}, \label{eq_arbitrary_init}
\end{align}

\noindent
where $\alpha_1 $ and $\beta_1 $ are real numbers and satisfy $\alpha_1 ^2 + \beta_1 ^2=1$. This can be interpreted as encoding two values, with amplitudes $\alpha_1 $ and $\beta_1 $ and phases $\varphi_1$ and $\phi_1$.

We apply a Hadamard gate, $H$, to this state, resulting in a superposition

\begin{align}
    H \ket{\psi} = H (\alpha_1  e^{i\varphi_1} \ket{0} + \beta_1  e^{i\phi_1} \ket{1}) &= \frac{1}{\sqrt{2}}[\alpha_1  e^{i\varphi_1} (\ket{0} + \ket{1}) + \beta_1  e^{i\phi_1} (\ket{0} - \ket{1})] \\
    &= \frac{1}{\sqrt{2}}[(\alpha_1  e^{i\varphi_1} + \beta_1  e^{i\phi_1})\ket{0} + (\alpha_1  e^{i\varphi_1} - \beta_1  e^{i\phi_1})\ket{1}].
\end{align}

\noindent
The probability of measuring $\ket{0}$ and $\ket{1}$ after the Hadamard gate application are

\begin{align}
    P(\ket{0}) &= |\bra 0 H \ket{\psi}|^2   \\
    &= \frac{1}{2} |\alpha_1  e^{i\varphi_1} + \beta_1  e^{i\phi_1}|^2 = \frac{1}{2} (1+ 2\alpha_1 \beta_1  \cos(\varphi_1 - \phi_1)),  \label{eq_prob_cos} \\
    P(\ket{1}) &= |\bra 1 H \ket{\psi}|^2  \\
    &= \frac{1}{2} |\alpha_1  e^{i\varphi_1} - \beta_1  e^{i\phi_1}|^2 = \frac{1}{2} (1 - 2\alpha_1 \beta_1  \cos(\varphi_1 - \phi_1)).
\end{align}

\noindent
The probability of measuring the state $\ket{0}$ directly corresponds to the interference term at the origin in the double-slit experiment, specifically the term $2 |A| |B| \cos(\varphi - \phi)$ in Eq.~\ref{eq_interference}. The probability $P(\ket{1})$ corresponds to destructive interference or dissimilarity between the two initial quantum states. The probabilities can also be written as

\begin{align}
    P(\ket{0}) &=\frac{1}{2} |\psi_1 + \psi_2|^2,\\
    P(\ket{1}) &=\frac{1}{2} |\psi_1 -  \psi_2|^2.
\end{align}

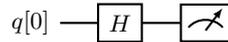
\begin{figure}[htpb]
\begin{center}
\begin{quantikz}
\lstick{$q[0]$} & \gate{H} & \meter{}
\end{quantikz}
\end{center}
  \caption{A single-qubit quantum circuit demonstrating superposition. The Hadamard gate creates an equal superposition of the $\ket{0}$ and $\ket{1}$ states, analogous to the superposition of waves in the double-slit experiment.}
\label{qc_circ1}
\end{figure}

\noindent
In the classical double-slit experiment, only $P(\ket{0})$ is directly visible. However, because the total probability of the waves must be 1 after passing through the wall, $P(\ket{1})$ must also be present after the wall. The quantum circuit diagram that models this double-slit phenomenon is shown in Figure~(\ref{qc_circ1}).

\noindent
The measurement of the qubit $q[0]$ collapses the quantum state and yields values 0 or 1, depending on the state of the system. By running the quantum circuit with multiple shots and summing over the results, we obtain probabilities that correspond to the values on the states. These probabilities reflect the "similarity" or "dissimilarity" between the initial states, providing a quantum-inspired way to assess similarity.

It is important to note that this quantum circuit does not directly compute the exact complex cosine similarity. Instead, it transforms the input state (encoding amplitudes and phases) and uses measurement probabilities as indicators of similarity. Higher probabilities for specific outcomes suggest greater alignment between the components of the input state. The results are statistical estimates and are subject to fluctuations. The values obtained are not numerically equal to the exact cosine cosine similarity but capture the concept of similarity in a different way, leveraging quantum superposition and interference. The encoding method, circuit design, and interpretation of probabilities are crucial for relating the quantum computation to semantic similarity. The potential quantum advantage lies in achieving more efficient or accurate similarity estimation compared to classical methods, particularly for high-dimensional data, although this remains a topic for further investigation.

\section{Quantum Circuit for Complex Cosine Similarity}

This section details the construction of a quantum circuit designed to estimate complex cosine similarity. Building upon the connection established in the previous section between the interference term and a simple quantum circuit, we now aim to connect complex cosine similarity to quantum systems. This requires: (1) identifying the necessary quantum circuit components for the measurement, and (2) deriving equations that relate the measurement results to the complex cosine similarity.

\subsection{Quantum Circuit Components for Similarity Measurement}

We begin by identifying the quantum circuit components required for the measurement. We start from the definition of complex cosine similarity in Eq.~(\ref{eq_cos_sim_complex})

\begin{align}
    {\bf S}_{C}({\bf a},{\bf b}) &= \frac{1}{||\mathbf{a}|| \ ||\mathbf{b}||} \sum_{i=1}^N a_{ci}^* b_{ci} \label{eq_cos_sim_complex10}
\end{align}

\noindent
We assume that vectors ${\bf a}$ and ${\bf b}$ are expressed as

\begin{align}
    {\bf a} &= [a_{c1}, a_{c2},...,a_{cN}] = [ a_1e^{i\varphi_1}, a_2e^{i\varphi_2} ,..., a_Ne^{i\varphi_N}  ].  \label{eq_vectora_real}\\
    {\bf b} &= [b_{c1}, b_{c2},...,b_{cN}] = [ b_1e^{i\phi_1}, b_2e^{i\phi_2},...,b_Ne^{i\phi_N}], \label{eq_vectorb_real}
\end{align}

\noindent
where $a_{i}$ and $b_{i}$ are real numbers. We also assume $L_2$ normalization of the vectors

\begin{align}
    ||\mathbf{a}||^2 &= \sum_{i=1}^Na_i^2  = 1, \label{eq_all_norm1}\\
    ||\mathbf{b}||^2 &= \sum_{i=1}^Nb_i^2  = 1. \label{eq_all_norm2}
\end{align}

\noindent
Expanding the equation, we obtain

\begin{align}
    {\bf S}_{C}({\bf a},{\bf b})
     &=  \sum_{i=1}^N(a_ib_i\cos(\phi_i-\varphi_i)) 
    +  ia_ib_i\sin(\phi_i-\varphi_i))  \label{eq_cos_sim_complex2}
\end{align}

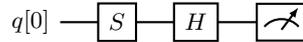
\begin{figure}[htpb]
\begin{center}
\begin{quantikz}
\lstick{$q[0]$} & \gate{S} & \gate{H} & \meter{}
\end{quantikz}
\end{center}
\caption{Quantum circuit demonstrating phase manipulation. The S-gate introduces a $\pi/2$ phase shift to the $\ket{1}$ state, enabling the estimation of both cosine and sine components of complex similarity.}
\label{qc_circ2}
\end{figure}

\noindent
We now consider quantum circuits capable of generating these terms. The circuit in Figure~(\ref{qc_circ1}) can be used to produce results for the cosine term. To generate the sine term, we observe that applying a $\pi/2$ rotation to one of the initial states yields the desired result. Therefore, we consider the circuit shown in Figure~(\ref{qc_circ2}). This circuit first applies the $S$-gate, which rotates the phase of state $\ket{1}$, followed by the Hadamard gate, producing a superposition between the states $\ket{0}$ and $\ket{1}$.

Using an arbitrary initial complex state as defined in Eq.~(\ref{eq_arbitrary_init}) and applying the steps in the circuit, we obtain the following state probabilities

\begin{align}
    P(\ket{0}) &=  \frac{1}{2} (1+ 2\alpha_1\beta_1 \sin(\varphi_1 - \phi_1)), \label{eq_prob_sin}\\
    P(\ket{1})     &=  \frac{1}{2} (1- 2\alpha_1\beta_1 \sin(\varphi_1 - \phi_1)).
\end{align}

\noindent
The probability $P(\ket{0})$ provides a quantum-inspired estimate related to the imaginary part of the component-wise complex cosine similarity, specifically the term involving $\sin(\varphi_1 - \phi_1)$. Note that

\begin{align}
    \sin(\phi_i-\varphi_i) = - \sin({\varphi_i-\phi_i}).
\end{align}

\noindent
Accurate calculations require careful attention to the signs of the phases. We now have the quantum circuit components, shown in Figures \ref{qc_circ1} and \ref{qc_circ2}, from which we can construct a circuit to estimate the complex cosine similarity.

\subsection{Derivation of the Complex Cosine Similarity Equation}

Our next aim is to derive an equation for the complex cosine similarity, ${\bf S}_C$, that can be effectively implemented using quantum circuit measurements. 
We start from Eq.~(\ref{eq_cos_sim_complex2})

\begin{align}
    {\bf S}_{C}({\bf a},{\bf b})
     &=  \sum_{i=1}^N a_i b_i e^{i(\phi_i - \varphi_i)} = \sum_{i=1}^N a_i b_i [\cos(\phi_i-\varphi_i) + i\sin(\phi_i-\varphi_i)]. \label{eq_cos_sim_complex2_rev}
\end{align}

\noindent
The key challenge lies in mapping these complex vector components to qubit states. Qubit states are represented by amplitudes, and these amplitudes must satisfy the normalization condition (sum of squares equals 1). To achieve this, we introduce a scaling factor, $c_i$, for each dimension $i$ to ensure that the scaled values can represent valid qubit amplitudes.

Let's define scaled values $\alpha_i$ and $\beta_i$ as

\begin{align}
    a_i &= c_i \alpha_i \\
    b_i &=  c_i \beta_i
\end{align}

\noindent
The crucial requirement is that for each dimension $i$, these scaled values satisfy the qubit normalization condition

\begin{align}
    \alpha_i^2 + \beta_i^2 = 1. \label{eq_qubit_norm}
\end{align}

\noindent
This ensures that $\alpha_i$ and $\beta_i$ can be interpreted as the amplitudes of a qubit state. Substituting the definitions of $\alpha_i$ and $\beta_i$, we get

\begin{align}
    c_i^2 = a_i^2 + b_i^2 \label{eq_scaling_factor}
\end{align}

\noindent
Using the results from Eqs.~(\ref{eq_prob_cos}) and (\ref{eq_prob_sin}), we can express the cosine and sine terms in terms of the probabilities obtained from quantum circuit measurements. Specifically, we can rewrite Eq.~(\ref{eq_prob_cos}) as

\begin{align}
    P(\ket{0})_i
     &=  \frac{1}{2}(1+2\alpha_i\beta_i\cos(\varphi_i-\phi_i)),
\end{align}

\noindent
from which

\begin{align}
    a_ib_i\cos(\varphi_i-\phi_i)
     &=  \frac{c_i^2}{2}(2P(\ket{0})_i-1) .
\end{align}

\noindent
Similarly, we can derive an equation for the sine term, and ${\bf S}_c$ becomes

\begin{align}
    {\bf S}_{C}({\bf a},{\bf b})
     &=  \frac{1}{2}\sum_{i=1}^N c_i^2(2P(\ket{0})_{2i-2}-1)
    -  \frac{i}{2}\sum_{i=1}^{N} c_i^2(2P(\ket{0})_{2i-1}-1), \label{eq_cos_sim_complex8_rev}
\end{align}

\noindent
where $P(\ket{0})_{2i-2}$ and $P(\ket{0})_{2i-1}$ are the probabilities of measuring the $\ket{0}$ state from the quantum circuits in Figures \ref{qc_circ1} and \ref{qc_circ2} for dimension $i$, respectively. 
Using quantum circuit controlled rotations, or multiple quantum circuits, one could potentially eliminate $c_i$ from the right-hand side of Eq.~(\ref{eq_cos_sim_complex8_rev}), but we leave it as is.

The magnitude of the complex cosine similarity can be calculated as

\begin{align}
    |{\bf S}_{C}({\bf a},{\bf b})|
     &=  \sqrt{\left[ \frac{1}{2}\sum_{i=1}^N c_i^2(2P(\ket{0})_{2i-2}-1) \right]^2
    +\left[ \frac{1}{2}\sum_{i=1}^N c_i^2(2P(\ket{0})_{2i-1}-1) \right]^2} \label{eq_cos_sim_complex_mag_rev}
\end{align}

\noindent
This equation provides a quantum-inspired method for estimating the complex cosine similarity between two vectors, leveraging the principles of quantum superposition and interference.

\subsection{Density Matrix Interpretation and Semantic Feature Analysis}

Equation~(\ref{eq_cos_sim_complex8_rev}) demonstrates that the coefficients $c_i^2$ are not arbitrary. Let's use the matrix formulation to write the same

\begin{align}
    {\bf S}_{C}({\bf a},{\bf b})
     &= 
\frac{1}{2}
\begin{bmatrix}
 c_1^2 & 0 & 0 & 0 & 0 \\
 0 & c_2^2 & 0 & 0 & 0 \\
 0 & 0 & \ddots & 0 & 0 \\
 0 & 0 & 0 & \ddots & 0 \\
 0 & 0 & 0 & 0 & c_N^2
\end{bmatrix}
\begin{bmatrix}
2P(\ket{0})_0-1\\
2P(\ket{0})_2-1 \\
 \vdots\\
 \vdots \\
 2P(\ket{0})_{2N-2}-1
\end{bmatrix}
-i
\frac{1}{2}
\begin{bmatrix}
 c_1^2 & 0 & 0 & 0 & 0 \\
 0 & c_2^2 & 0 & 0 & 0 \\
 0 & 0 & \ddots & 0 & 0 \\
 0 & 0 & 0 & \ddots & 0 \\
 0 & 0 & 0 & 0 & c_N^2
\end{bmatrix}
\begin{bmatrix}
2P(\ket{0})_1-1\\
2P(\ket{0})_3-1 \\
 \vdots\\
 \vdots \\
 2P(\ket{0})_{2N-1}-1
\end{bmatrix} \\
&=
\boldsymbol{\rho}_c\mathbf{P}_c -i\boldsymbol{\rho}_c\mathbf{P}_i
\end{align}

\noindent
where

\begin{align}
    \boldsymbol{\rho}_c
     &= 
\frac{1}{2}
\begin{bmatrix}
 c_1^2 & 0 & 0 & 0 & 0 \\
 0 & c_2^2 & 0 & 0 & 0 \\
 0 & 0 & \ddots & 0 & 0 \\
 0 & 0 & 0 & \ddots & 0 \\
 0 & 0 & 0 & 0 & c_N^2
\end{bmatrix}
\end{align}

\noindent
The matrix $\boldsymbol{\rho}_c$ can be interpreted as a quantum density matrix, describing the statistical state of a quantum system. A key property of density matrices is that their trace is equal to 1

\begin{align}
    \text{Tr}(\boldsymbol{\rho}_c)
     &= 1.
\end{align}

\noindent
In this context, $\boldsymbol{\rho}_c$ represents the uncertainty in our knowledge of the embedding vectors. Each diagonal element, $c_i^2 / 2$, weights the contribution of the $i$-th dimension to the similarity calculation, reflecting the combined magnitude ($a_i^2 + b_i^2$) of the corresponding components in vectors {\bf a} and {\bf b}. The diagonal form of the density matrix indicates that the dimensions are treated as statistically independent in the component-wise similarity calculation.

If the original embedding vectors {\bf a} and {\bf b} are L2 normalized, then

\begin{align}
    \text{Tr}(\boldsymbol{\rho}_c)
     &= \frac{1}{2}\sum_{i=1}^N c_i^2 = \frac{1}{2}\sum_{i=1}^N (a_i^2 + b_i^2) =  \frac{1}{2}\sum_{i=1}^N a_i^2 +  \frac{1}{2}\sum_{i=1}^N b_i^2 = 1.
\end{align}

\noindent
This constraint, imposed by the properties of density matrices and L2 normalization, suggests that the individual components $a_i^2$ and $b_i^2$ might also have a specific meaning beyond simply being squared components of a vector.  Specifically, if we interpret the $a_i^2$ values as the diagonal elements (and therefore eigenvalues) of a matrix representing the embedding vector {\bf a} in semantic space, then it suggests that each dimension of the embedding space contributes a specific, quantifiable amount to the overall semantic representation of the word.

Based on quantum density matrix definitions, many established results from quantum mechanics become applicable, \cite{ref_merzbacher}. In quantum mechanics, an observable represents a physical quantity that can be measured, while its expectation value is the average value one would obtain from many measurements on identically prepared systems.
If $ {\bf S}_{C}$ is an observable of the system, then its expectation value is given by

\begin{align}
 \langle {\bf S}_{C} \rangle &= \text{Tr}(\boldsymbol{\rho}_c{\bf S}_{C})
\end{align}

\noindent
with the spectral resolution

\begin{align}
 {\bf S}_{C} &= \sum_{i=1}^N c_i^2P_i
\end{align}

\noindent
where $P_i$ is the projection operator onto the eigenspace corresponding to eigenvalue $c_i^2$. Analogously, the vectors
\begin{equation*}
\begin{bmatrix}
2P(\ket{0})_0-1\\
2P(\ket{0})_2-1 \\
 \vdots\\
 \vdots \\
 2P(\ket{0})_{2N-2}-1
\end{bmatrix}
\text{ and }
\begin{bmatrix}
2P(\ket{0})_1-1\\
2P(\ket{0})_3-1 \\
 \vdots\\
 \vdots \\
 2P(\ket{0})_{2N-1}-1
\end{bmatrix}
\end{equation*}
can be interpreted as projection operators in the eigenspace of the qubits.

The observable ${\bf S}_{C}$ has eigenvalues $c_i^2$, representing the contribution of each dimension to the complex cosine similarity. The density matrix $\boldsymbol{\rho}_c$, on the other hand, also possesses eigenvalues. These eigenvalues of $\boldsymbol{\rho}_c$ represent the probabilities of finding the system in a particular eigenstate, which can be interpreted as the prominence of specific semantic features within the embedded words. The magnitude of each eigenvalue reflects the importance of the corresponding semantic feature in characterizing the words. The eigenvectors of $\boldsymbol{\rho}_c$ provide a basis for the semantic space, with each eigenvector corresponding to a direction or axis, and its components indicating the contribution of each dimension to that axis.
A detailed exploration of these consequences will be addressed in future work.

\section{Illustrative Example: Semantic Similarity between "Dog" and "Cat"}

This section provides an illustrative example of calculating the semantic similarity between the words "dog" and "cat" using the quantum circuit approach described above. To simplify the demonstration, we represent each word using a two-dimensional complex embedding. The embedding vectors are defined as

\begin{align}
    {\bf E}_1(\text{dog})&=  [a_{1} e^{i\varphi_1} , a_{2} e^{i\varphi_2}] , \\
    {\bf E}_2(\text{cat}) &= [b_{1} e^{i\phi_1} , b_{2} e^{i\phi_2} ].
\end{align}

\noindent
The embedding vectors are normalized such that $|{\bf E}_i|^2= 1$. The complex cosine similarity is then given by

\begin{align}
    {\bf S}_{C}({\bf a},{\bf b})
     &=  a_1b_1\cos(\phi_1-\varphi_1) +   ia_1b_1\sin(\phi_1-\varphi_1)+ a_2b_2\cos(\phi_2-\varphi_2)
    +ia_2b_2\sin(\phi_2-\varphi_2).  \label{eq_cos_sim_complex7}
\end{align}

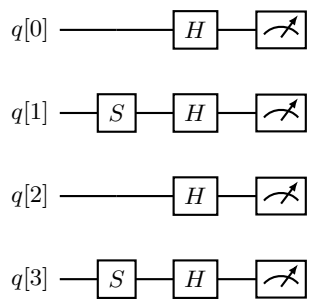
\begin{figure}[htpb]
\begin{center}
\begin{quantikz}
\lstick{$q[0]$} & \qw & \gate{H} & \meter{} & \\
\lstick{$q[1]$} & \gate{S} & \gate{H} & \meter{} \\
\lstick{$q[2]$} & \qw & \gate{H} & \meter{} & \\
\lstick{$q[3]$} & \gate{S} & \gate{H} & \meter{}
\end{quantikz}
\end{center}
\caption{Quantum circuit for calculating the complex cosine similarity between simplified "dog" and "cat" embeddings. Qubits 0 and 2 estimate the real components, while qubits 1 and 3, incorporating S-gates, estimate the imaginary components.}
\label{qc_circ3}
\end{figure}

\noindent
To calculate the complex cosine similarity, we require a quantum circuit to estimate each term in the equation. Several circuit architectures are possible. We could use a single qubit and execute the quantum job four times, use two qubits and execute separate jobs for the real and imaginary parts, or use four qubits to calculate all values in a single job. For this example, we select the four-qubit approach. The quantum circuit is shown in Figure~\ref{qc_circ3}. The qubits are organized to correspond to the terms in Eq.~(\ref{eq_cos_sim_complex7}).

\noindent
In this circuit, the first qubit estimates the real part of the first term in Eq.~(\ref{eq_cos_sim_complex7}), and the second qubit estimates the imaginary part, and so on. The qubits are independent and do not interact with each other. For larger embeddings with $2N$ dimensions, the circuit can be constructed using $2N$ qubits, following the same logic.

The initial states for the quantum circuit are

\begin{align}
\ket{q[0]}_{initial} &= \alpha_1e^{i\varphi_1}\ket{0}  + \beta_1e^{i\phi_1}\ket{1} \label{eq_init1}\\
\ket{q[1]}_{initial} &= \alpha_1e^{i\varphi_1}\ket{0}  + \beta_1e^{i\phi_1}\ket{1} \\
\ket{q[2]}_{initial} &= \alpha_2e^{i\varphi_2}\ket{0}  + \beta_2e^{i\phi_2}\ket{1} \\
\ket{q[3]}_{initial} &= \alpha_2e^{i\varphi_2}\ket{0}  + \beta_2e^{i\phi_2}\ket{1} \label{eq_init4}
\end{align}

\noindent
where $\alpha_i$ and $\beta_i$ are real numbers that satisfy $\alpha_i^2+\beta_i^2=1$. We encode the word embeddings into the quantum circuit. After applying the phase and Hadamard gates, we obtain

\begin{align}
\ket{q[0]} &= \frac{1}{\sqrt{2}}\left[(\alpha_1e^{i\varphi_1} + \beta_1e^{i\phi_1})\ket{0} + (\alpha_1e^{i\varphi_1} - \beta_1e^{i\phi_1})\ket{1}\right] \\
\ket{q[1]} &= \frac{1}{\sqrt{2}}\left[(\alpha_1e^{i\varphi_1} + i\beta_1e^{i\phi_1})\ket{0} + (\alpha_1e^{i\varphi_1} - i\beta_1e^{i\phi_1})\ket{1}\right] \\
\ket{q[2]} &= \frac{1}{\sqrt{2}}\left[(\alpha_2e^{i\varphi_2} + \beta_2e^{i\phi_2})\ket{0} + (\alpha_2e^{i\varphi_2} - \beta_2e^{i\phi_2})\ket{1}\right] \\
\ket{q[3]} &= \frac{1}{\sqrt{2}}\left[(\alpha_2e^{i\varphi_2} + i\beta_2e^{i\phi_2})\ket{0} + (\alpha_2e^{i\varphi_2} - i\beta_2e^{i\phi_2})\ket{1}\right]
\end{align}

\noindent
The final state is a tensor product of these qubit states

\begin{equation}
	\ket{\psi}_{final} = \ket{q[0]} \otimes \ket{q[1]} \otimes \ket{q[2]} \otimes \ket{q[3]}.
\end{equation}

\noindent
The probabilities of measuring the qubit states after a single shot are

\begin{align}
P(q[0] \rightarrow 0) &= \frac{1}{2}(1+ 2\alpha_1\beta_1\cos(\varphi_1 - \phi_1)) \label{eq_prob1_final}\\
P(q[0] \rightarrow 1) &= \frac{1}{2}(1 - 2\alpha_1\beta_1\cos(\varphi_1 - \phi_1)) \\
P(q[1] \rightarrow 0) &= \frac{1}{2}(1 + 2\alpha_1\beta_1\sin(\varphi_1 - \phi_1)) \\
P(q[1] \rightarrow 1) &= \frac{1}{2}(1- 2\alpha_1\beta_1\sin(\varphi_1 - \phi_1)) \\
P(q[2] \rightarrow 0) &= \frac{1}{2}(1+ 2\alpha_2\beta_2\cos(\varphi_2 - \phi_2)) \\
P(q[2] \rightarrow 1) &= \frac{1}{2}(1- 2\alpha_2\beta_2\cos(\varphi_2 - \phi_2)) \\
P(q[3] \rightarrow 0) &= \frac{1}{2}(1+ 2\alpha_2\beta_2\sin(\varphi_2 - \phi_2)) \\
P(q[3] \rightarrow 1) &= \frac{1}{2}(1- 2\alpha_2\beta_2\sin(\varphi_2 - \phi_2))\label{eq_prob8_final}
\end{align}

\noindent
We use Qiskit and the quantum Aer-simulator to calculate the results. Qiskit is an open-source Python SDK for quantum computing that enables users to design, simulate, and execute quantum algorithms on both simulators and real quantum hardware. It offers a modular approach with tools for quantum algorithm development, simulation, error correction, and applications in fields like chemistry, machine learning, finance, and optimization.

To demonstrate the approach, we assume an embedding model produces the following vectors

\begin{align}
    {\bf E}_1(\text{dog})&= [0.4 e^{i\pi/6}, \sqrt{1-0.4^2}e^{i\pi/2}], \\
    {\bf E}_2(\text{cat}) &=  [0.5 e^{i\pi/4}, \sqrt{1-0.5^2}e^{i\pi/3}] .
\end{align}

\noindent
These vectors have been selected to be arbitrary, solely to illustrate the approach and assess its accuracy. In the next section, we present a more realistic case. 
We shoot the words to the double-slit experiment.
Applying Qiskit, the simulator provides estimates for the probabilities in Eqs.~(\ref{eq_prob1_final})-(\ref{eq_prob8_final}), given the initial conditions in Eqs.~(\ref{eq_init1})-(\ref{eq_init4}).

\noindent
We have calculated the exact final probabilities of the qubit states using Eqs.~(\ref{eq_prob1_final})-(\ref{eq_prob8_final}). The results are shown in Table~\ref{tab_results1}, along with the experimental results obtained using the Qiskit quantum simulator. We used 10000 shots for the simulation.

\begin{table}[htpb]
\centering
\caption{Measurement probabilities obtained from theoretical calculations and a quantum simulator for the "dog" and "cat" example, demonstrating the ability of quantum simulation to approximate the expected quantum behavior.}
\begin{tabular}{c|c|c}
\toprule
                       & Exact results & Simulation \\
$q[0]$, state 0  & 0.971 & 0.973  \\
$q[0]$, state 1  & 0.029 & 0.027 \\
$q[1]$, state 0 & 0.374 & 0.371 \\
$q[1]$, state 1 & 0.626 & 0.629 \\
$q[2]$, state 0 & 0.932 & 0.936 \\
$q[2]$, state 1 & 0.068& 0.064 \\
$q[3]$, state 0 & 0.750 & 0.751 \\
$q[3]$, state 1 & 0.250 & 0.249 \\
\end{tabular}
\label{tab_results1}
\end{table}

The complex cosine similarity is calculated as

\begin{align}
    {\bf S}_{C}({\bf a},{\bf b})
     &=  \frac{1}{2}\left [ c_1^2(2P(q[0] \rightarrow 0)-1) + c_2^2(2P(q[2] \rightarrow 0)-1) \right ] \nonumber \\
    &-  \frac{i}{2}\left [ c_1^2(2P(q[1] \rightarrow 0)-1) + c_2^2(2P(q[3] \rightarrow 0)-1) \right ]. \label{eq_cos_sim_complex8_rev5}
\end{align}

\noindent
Using the exact probabilities, the complex similarity is

\begin{align}
   {\bf S}_C(\text{exact}) = 0.881-i0.345
\end{align}

\noindent
Using the probabilities obtained from the simulation, the complex similarity is

\begin{align}
   \langle {\bf S}_C(\text{simulated}) \rangle = 0.886-i0.346
\end{align}

\noindent
The results from the simulation are in close agreement with the exact calculation. The exact results are based on direct calculation using the equations, while the simulated results are obtained from the quantum computer simulator, which provides probabilities and statistical averages.

Depending on the specific use case, the simulated result may be sufficiently accurate. For example, cosine similarity is often used to retrieve the top k best matches in RAG systems, where high precision of ${\bf S}_C$ is not necessary required. The accuracy also depends on the corpus used to train the embedding model. If the training corpus is large, the embedding space is dense, meaning there are many states that are close to each other. In such cases, small differences in the decimal places of the cosine similarity can be meaningful.

\section{Quantum Computation of Cosine Similarity with a Real LLM}

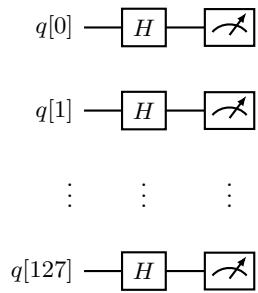
\begin{figure}[htpb]
\begin{center}
\begin{quantikz}
\lstick{$q[0]$} & \gate{H} & \meter{} & \\
\lstick{$q[1]$} & \gate{H} & \meter{} \\
\lstick{\vdots} & \vdots & \vdots & \\
\lstick{$q[127]$} & \gate{H} & \meter{}
\end{quantikz}
\end{center}
   \caption{Quantum circuit used to calculate cosine similarity between embeddings generated by Google's Sentence Transformer. The 128 qubits represent the 128 dimensions of the MRL-truncated embedding vectors.}
\label{qc_circ4}
\end{figure}

This section presents the results of computing cosine similarity between embedding vectors generated by a real-world LLM on a real quantum computer. We used Google's Sentence Transformer, `google/embedding-gemma-300m` (available on Hugging Face), to generate the embedding vectors. EmbeddingGemma, a 300M parameter model, leverages Gemma 3 (with T5Gemma initialization) and the technology used in Gemini models to produce vector representations of text suitable for tasks like search, retrieval, classification, clustering, and semantic similarity search. Trained on data in over 100 languages, EmbeddingGemma outputs vectors with a default dimension of 768. This model was selected as a representative example of a modern, accurate, and widely used sentence embedding model. The specific choice of model is not critical to the demonstration of the quantum computation; any comparable model producing suitable embedding vectors could be used. To reduce the qubit requirements for this demonstration, we employed Matryoshka Representation Learning (MRL) to truncate the output to 128 dimensions. MRL was chosen to reduce the dimensionality of the embeddings to match the qubit constraints of the available quantum hardware. The quantum circuit used is shown in Figure~\ref{qc_circ4}.

The quantum computations were performed using the IBM Quantum Cloud, which provides access to IBM's quantum computing systems and resources. This platform allows researchers, developers, and educators to explore and advance quantum computing without the need for specialized hardware. It offers access to quantum processors, the Qiskit software development kit, and a collaborative environment. 

As a concrete example, we calculated the cosine similarity between the following sentences:
\begin{equation*}
\text{``the quick brown fox jumps over the lazy dog''}
\end{equation*}
and
\begin{equation*}
\text{``the brown dog jumps over the lazy cat''}
\end{equation*}
These sentences were chosen to illustrate the model's ability to capture semantic similarity despite variations in wording. The Google Sentence Transformer generated 128-dimensional embedding vectors for these sentences (after MRL projection). These vectors were then normalized to unit length, a standard procedure before calculating cosine similarity. Calculating the cosine similarity required 128 qubits. The initial state was prepared using amplitude encoding, and a quantum circuit with Hadamard gates was employed. The Python program, developed using Qiskit, was executed on the IBM Quantum Cloud, utilizing 4096 shots with the \texttt{ibm\_fez} backend. The choice of the backend was based on its availability and sufficient qubit count at the time of execution. The resulting quantum computation yielded a cosine similarity of

\begin{align}
  \langle {\bf S}_C(\text{quantum}) \rangle &= 0.8635
\end{align}

\noindent
The corresponding theoretical result, calculated classically, is

\begin{align}
   {\bf S}_C(\text{exact}) = 0.8682
\end{align}

\noindent
The cosine similarity is a real number in this case due to the use of real-valued embeddings. The results demonstrate reasonable agreement between the quantum computation and the classical calculation, suggesting that the quantum circuit is capturing the essential features of the semantic similarity between the sentences. However, a small discrepancy exists between the quantum and classical results.  It's important to acknowledge that current quantum hardware is susceptible to noise and errors, which can impact the accuracy of computations.  

\subsection{Addressing Noise and Error in Quantum Computations}

The experimental validation of our quantum-inspired approach to LLM embedding analysis, while promising, is inherently limited by the presence of noise and errors in current quantum hardware. These imperfections can significantly impact the accuracy and reliability of quantum computations, leading to deviations from the theoretically expected results. Several factors contribute to these errors, including decoherence, where qubit coherence times are finite, meaning that the quantum state of a qubit can degrade over time due to interactions with the environment; gate errors, where quantum gates are not perfect and can introduce errors during their execution; measurement errors, where the measurement process itself can introduce errors; and crosstalk, where unintended interactions between qubits can lead to errors. The overall fidelity of the quantum computation, a measure of how closely the actual quantum state matches the ideal state, is thus reduced by these imperfections.

To mitigate the effect of noise and errors, several techniques can be employed. Error mitigation techniques aim to reduce the impact of errors without requiring full quantum error correction. Examples include zero-noise extrapolation (ZNE), probabilistic error cancellation (PEC), and readout error mitigation. Quantum error correction (QEC) involves encoding quantum information in a redundant manner, allowing for the detection and correction of errors. Advancements in quantum hardware, such as longer coherence times, more accurate gates, and reduced crosstalk, will also contribute to reducing noise and errors in quantum computations.

Future research should focus on implementing and evaluating these error mitigation techniques in the context of quantum-enhanced LLM embedding analysis, and rigorously assessing the statistical significance of the differences between quantum and classical results, accounting for noise and error. Furthermore, as quantum hardware continues to improve, the impact of noise and errors will be reduced, making quantum computations more accurate and reliable.

\subsection{Code Listing}

This subsection shows the Python code that was used in quantum computer calculation.

\begin{lstlisting}[
    language=Python,
    basicstyle=\ttfamily\footnotesize,
    numbers=left,
    numberstyle=\tiny\color{gray},
    stepnumber=1,
    numbersep=5pt,
    backgroundcolor=\color{white},
    showspaces=false,
    caption={Qiskit code for quantum cosine similarity calculation.},
    label=lst:quantum_code,
    breaklines=true
]
import numpy as np
import pickle
from qiskit import QuantumCircuit, transpile
from qiskit_ibm_runtime import QiskitRuntimeService, Session, Sampler

def run_and_store_results(vectora_q, vectorb_q, filename='quantum_results.pkl'):

    if len(vectora_q) != len(vectorb_q):
        raise ValueError("Vectors must have the same length.")

    num_qubits = len(vectora_q)
    qc = QuantumCircuit(num_qubits, num_qubits)

    for i in range(num_qubits):
        a = vectora_q[i]
        b = vectorb_q[i]

        initial_state = np.array([a, b], dtype=np.complex128)
        norm = np.linalg.norm(initial_state)
        normalized_state = initial_state / norm

        qc.initialize(normalized_state, i)

    qc.h(range(num_qubits))
    qc.measure(range(num_qubits), range(num_qubits))

    try:
        service = QiskitRuntimeService()

        backend = service.least_busy(simulator=False, operational=True, min_num_qubits=num_qubits)
        print(f"Using backend: {backend.name}")

        transpiled_qc = transpile(qc, backend)

        with Session(backend=backend) as session:
            sampler = Sampler()

            job = sampler.run([transpiled_qc], shots=4096)
            print(f"Job ID: {job.job_id()}")
            result = job.result()

            with open(filename, 'wb') as f:
                pickle.dump(result, f)

            print(f"Results stored in {filename}")

    except Exception as e:
        print(f"Error running the circuit: {e}")
        print(f"Error type: {type(e)}")
        print(f"Error message: {e}")
\end{lstlisting}

\subsection{Explanation}

The \texttt{run\_and\_store\_results} function takes two vectors, \texttt{vectora\_q} and \texttt{vectorb\_q}, as input, representing the amplitudes used to initialize the qubits. The function begins by checking if the input vectors have the same length. It then creates a \texttt{QuantumCircuit} with the number of qubits equal to the length of the input vectors. For each qubit, the state is initialized using the corresponding amplitudes from \texttt{vectora\_q} and \texttt{vectorb\_q} through amplitude encoding, where the values from the vectors are directly used to define the initial quantum state, normalizing the initial state to ensure it represents a valid quantum state. Subsequently, a Hadamard gate (\texttt{qc.h}) is applied to each qubit, creating a superposition essential for leveraging quantum interference to estimate the cosine similarity, and all qubits are measured. The function then connects to the IBM Quantum Cloud using \texttt{QiskitRuntimeService()}, selects a real quantum backend using \texttt{service.least\_busy()}, and transpiles the circuit for the selected backend using \texttt{transpile()}. A session is created with the backend, and a \texttt{Sampler} instance is created. The circuit is run using the \texttt{Sampler} within the session, which efficiently estimates probabilities without returning the full state vector. Finally, the results are stored in a pickle file. A try-except block is included to catch any errors during the quantum circuit execution, printing the error type and message if an error occurs. The key aspects of this implementation include the use of amplitude encoding for qubit initialization, the application of the Hadamard gate to leverage quantum interference, and the use of the Sampler primitive for efficient probability estimation.

\section{Complex Embeddings and the Full Quantum Mechanical Framework}

Given that LLM embeddings can be real-valued, one might question the necessity of quantum mechanics, which relies on complex values.
This section explores the potential for leveraging complex embeddings to more fully utilize the quantum mechanical framework when calculating semantic similarity. 
A 2N-dimensional real vector can be represented as an N-dimensional complex vector using the following transformations:

\begin{align}
    \alpha_i &= \sqrt{a_{2i-1}^2+a_{2i}^2} \\
    \varphi_i &= \begin{cases}
               \arccos(a_{2i-1}/\alpha_i) & \text{if } a_{2i} \geq 0 \\
               -\arccos(a_{2i-1}/\alpha_i) & \text{if } a_{2i} < 0
             \end{cases} \\
    a_{2-i1} &= \alpha_i \cos(\varphi_i) \\
    a_{2i} &= \alpha_i \sin(\varphi_i)
\end{align}

\noindent
While this representation allows a more compact encoding, the computational cost on a quantum computer is contingent on the embedding type.

Consider complex embeddings with real coefficients, ${\bf a} = [a_1 e^{i\varphi_1}, ..., a_N e^{i\varphi_N}]$ and ${\bf b} = [b_1 e^{i\phi_1}, ..., b_N e^{i\phi_N}]$. The complex cosine similarity can be expressed as

\begin{align}
    {\bf S}_{C}({\bf a},{\bf b})
     &= \sum_{i=1}^N a_i b_i e^{i(\phi_i - \varphi_i)}
\end{align}

\noindent
We can factor out the first term

\begin{align}
    {\bf S}_{C}({\bf a},{\bf b})
     &=  a_1 b_1 e^{i(\phi_1 - \varphi_1)} + \sum_{i=2}^N a_i b_i e^{i(\phi_i - \varphi_i)} \\
     &=  e^{i(\phi_1 - \varphi_1)} \left[ a_1 b_1 + \sum_{i=2}^N a_i b_i e^{i(\phi_i - \varphi_i) - i(\phi_1 - \varphi_1)} \right].
\end{align}

\noindent
Let's define ${\bf \tilde{S}}_{C}({\bf \tilde{a}},{\bf \tilde{b}})$ as

\begin{align}
    {\bf \tilde{S}}_{C}({\bf \tilde{a}},{\bf \tilde{b}}) &=  a_1 b_1 + \sum_{i=2}^N a_i b_i e^{i(\phi_i - \varphi_i) - i(\phi_1 - \varphi_1)}
\end{align}

\noindent
then

\begin{align}
    {\bf S}_{C}({\bf a},{\bf b}) = e^{i(\phi_1 - \varphi_1)} {\bf \tilde{S}}_{C}({\bf \tilde{a}},{\bf \tilde{b}}).
\end{align}

\noindent
Taking the magnitude gives

\begin{align}
    |{\bf S}_{C}({\bf a},{\bf b})| = |{\bf \tilde{S}}_{C}({\bf \tilde{a}},{\bf \tilde{b}})|.
\end{align}

\noindent
This allows us to focus on calculating $|{\bf \tilde{S}}_{C}({\bf \tilde{a}},{\bf \tilde{b}})|$. By calculating $a_1b_1$ using one qubit, we can, in principle, reduce the total qubit count requirement by one. However, the primary significance of this transformation lies not in the qubit reduction itself, but in the fact that it allows us to operate within the full quantum mechanical framework. Even starting with real embeddings, by transforming to a complex representation, we gain access to the full power of quantum mechanics, including the ability to manipulate phases and exploit interference effects.

When using a complex representation, if the input vectors are real, the resulting cosine similarity will also be a real number.
The imaginary component can be viewed as a helper dimension. It provides additional degrees of freedom that can be exploited within a quantum circuit. It allows for operations like phase flips and rotations that can manipulate the quantum state in ways not possible with purely real-valued representations. This added flexibility could potentially lead to the design of more efficient and tailored quantum circuits for semantic similarity calculations, even if the initial qubit savings are minimal.

While this initial approach focuses on utilizing the full quantum mechanical framework, significant qubit reduction ultimately requires more sophisticated techniques that exploit the underlying structure of the problem. Several techniques can be employed to achieve qubit reduction. Exploiting sparsity, if the embedding vectors exhibit a high degree of sparsity, techniques like compressed sensing or sparse quantum algorithms could substantially reduce the qubit requirement. The potential reduction depends on the degree of sparsity in the embedding vectors. Quantum Phase Estimation (QPE) offers another approach, providing a way to estimate eigenvalues related to cosine similarity. The number of qubits required for QPE scales logarithmically with the desired precision, potentially offering significant savings compared to direct amplitude encoding. Variational Quantum Algorithms (VQAs) present a third possibility, as they can be designed to directly learn the cosine similarity between embedding vectors. The qubit requirement for VQAs depends on the complexity of the chosen ansatz (parameterized quantum circuit). Quantum Singular Value Decomposition (QSVD) can decompose the matrix formed by the embedding vectors, allowing for cosine similarity estimation based on singular values. The qubit requirement for QSVD depends on the size of the matrix being decomposed.

\section{Potential Applications and Future Research Directions}

The presented cosine similarity calculation using a quantum computer alone is not currently practical, as classical methods are faster for real-valued embeddings. However, the quantum-inspired approach presented in this paper could become relevant as part of a larger quantum-classical process, such as training sentence transformers.

Sentence transformers generate dense vector representations for sentences, capturing semantic meaning. They are fine-tuned on large datasets to outperform generic transformer models on tasks like semantic search and text classification. Training involves optimizing sentence embeddings using sentence pairs or triplets and evaluating performance on downstream tasks. Quantum algorithms could accelerate the training process by efficiently optimizing the embedding space or providing better initial parameter estimates. Quantum optimization algorithms could be used to find the optimal parameters for the sentence transformer model. This is a relatively unexplored area.

Other potential applications and future research directions include:
Implicit feature maps and quantum kernels for enhanced representation. If the qubit amplitudes are the result of a complex feature map, a quantum computer could implement this map implicitly using a quantum kernel. This could lead to a more efficient feature representation, as quantum kernels can perform computations in high-dimensional spaces intractable for classical computers. Variational Quantum Eigensolver (VQE) or Quantum Approximate Optimization Algorithm (QAOA) could be used to find the optimal feature map. Designing effective quantum kernels remains a challenge.

Entanglement for high-dimensional semantic analysis is another possibility. Quantum computers can create and manipulate entangled states in high-dimensional Hilbert spaces, potentially allowing cosine similarity calculations between complex quantum states that are impossible to process classically. The Hilbert space of N qubits has dimension $2^N$. Building and controlling large numbers of qubits is a major technological challenge.

Quantum amplitude estimation for high-precision similarity measurement is also promising. If high precision is required, Quantum Amplitude Estimation (QAE) can estimate probabilities with a speedup compared to classical Monte Carlo methods. QAE uses Grover's search algorithm to estimate the amplitude of a quantum state. QAE requires a relatively large number of qubits and coherent gate operations.

Hybrid quantum-classical approaches for semantic analysis offer a promising direction. Combining the strengths of both quantum and classical computation, hybrid algorithms could leverage quantum computers for specific tasks, such as similarity calculations or feature extraction. Designing effective hybrid algorithms requires careful consideration of the strengths and weaknesses of both quantum and classical computation.

\section{Discussion}

This paper has explored the application of quantum formalism to the analysis of LLM embeddings, providing theoretical insights and experimental validation of a connection between these fields. We demonstrated the estimation of cosine similarity between LLM embeddings using quantum circuits and a real quantum computer, suggesting a potential pathway for quantum-enhanced semantic analysis. Quantum algorithms, specifically designed for this task, could unlock advantages in discovering subtle semantic relationships or accelerating similarity calculations in high-dimensional spaces.

Our exploration suggests that LLM embedding spaces exhibit characteristics that resonate with quantum mechanics. The finite vocabulary of tokens hints at an underlying discreteness, aligning with the concept of quantization. By extending the embedding space to the complex domain and drawing parallels to the double-slit experiment, we modeled potential semantic interference effects. 

The quantum-inspired formalism offers a new perspective for analyzing LLM embeddings. The analogy to quantum mechanics suggests that concepts like superposition and interference may offer insights into semantic representation and processing. The use of quantum circuits provides experimental tools to explore these concepts.

Several limitations must be acknowledged. The current implementation relies on amplitude encoding, leading to a qubit requirement that scales linearly with embedding dimensionality. This presents a challenge for scaling to larger LLMs. Experiments on real quantum computers are susceptible to noise and limited qubit coherence times, affecting accuracy. The relatively small LLM and truncated embedding dimension also limit the generalizability of our findings.
Despite these limitations, this work suggests promising directions for future research. These include developing more efficient quantum algorithms for semantic similarity, such as those based on quantum phase estimation or variational quantum eigensolvers, and investigating quantum error correction techniques.

Beyond semantic similarity, this approach could potentially be extended to other NLP tasks, such as text generation and machine translation, by modeling probabilistic relationships between words and phrases using quantum circuits. The density matrix representation of semantic information could also be used as a feature representation for downstream machine learning models.

This work highlights the potential for quantum computing to contribute to a deeper understanding of natural language and intelligence. By demonstrating a connection between semantic spaces and quantum mechanics, we hope to inspire further research into quantum-inspired models for cognitive science and artificial intelligence. Assuming that quantum mechanics provides a valid framework for LLM semantic spaces, and if quantum computing can be effectively applied to LLM training and predictions, it could significantly reduce the reliance on large GPU farms, potentially democratizing LLM development and usage by lowering the barrier to entry. 

While high accuracy is crucial for many LLM applications, current expectations often exceed the capabilities of these models, leading to disappointment. If LLMs indeed operate according to quantum mechanical principles, as suggested by our findings, then a degree of inherent uncertainty is unavoidable. However, a good understanding of quantum mechanics provides the tools to manage and control this uncertainty, potentially enabling developers to achieve more robust and reliable results from LLMs, even in the face of inherent probabilistic behavior.

In conclusion, this paper provides evidence for the potential of applying quantum formalism to the analysis of LLM embeddings. While challenges remain, the combination of theoretical insights and experimental validation suggests that quantum circuits may offer new tools for exploring these systems. Further research is needed to realize the potential of quantum natural language processing.

\end{document}